\documentclass[%
 aip,
 amsmath,amssymb,
 reprint,%
]{revtex4-1}

\usepackage{graphicx}
\usepackage{dcolumn}
\usepackage{bm}
\usepackage[utf8]{inputenc}
\usepackage[T1]{fontenc}
\usepackage{mathptmx}
\usepackage{etoolbox}
\usepackage{xcolor}

\makeatletter
\def\@email#1#2{%
 \endgroup
 \patchcmd{\titleblock@produce}
  {\frontmatter@RRAPformat}
  {\frontmatter@RRAPformat{\produce@RRAP{*#1\href{mailto:#2}{#2}}}\frontmatter@RRAPformat}
  {}{}
}%
\makeatother
\begin{document}

\preprint{AIP/123-QED}

\title{Spatial homogeneity of superconducting order parameter in NbN films grown by atomic layer deposition}

\author{J. Lorenz}
\affiliation{Peter Gr\"unberg Institute (PGI-3), Forschungszentrum J\"ulich, 52425 J\"ulich, Germany}
\affiliation{Institute of Experimental Physics IV A, RWTH Aachen Universität, 52074 Aachen, Germany}
\affiliation{J\"ulich Aachen Research Alliance (JARA), Fundamentals of Future Information Technology, 52425 J\"ulich, Germany}

\author{S. Linzen}%
 \email{sven.linzen@leibniz-ipht.de.}
 \affiliation{ Leibniz Institute of Photonic Technology, 07702 Jena, Germany}%
 
\author{M. Ziegler}%
 \affiliation{ Leibniz Institute of Photonic Technology, 07702 Jena, Germany}%
 
\author{G. Oelsner}%
 \affiliation{ Leibniz Institute of Photonic Technology, 07702 Jena, Germany}%
 
\author{R. Stolz}%
 \affiliation{ Leibniz Institute of Photonic Technology, 07702 Jena, Germany}%

\author{F. S. Tautz}%
\affiliation{Peter Gr\"unberg Institute (PGI-3), Forschungszentrum J\"ulich, 52425 J\"ulich, Germany}%
\affiliation{Institute of Experimental Physics IV A, RWTH Aachen Universität, 52074 Aachen, Germany}%
\affiliation{J\"ulich Aachen Research Alliance (JARA), Fundamental of Future Information Technology, 52425 J\"ulich, Germany}%

\author{F. Lüpke}%
\email{f.luepke@fz-juelich.de}%
\affiliation{Peter Gr\"unberg Institute (PGI-3), Forschungszentrum J\"ulich, 52425 J\"ulich, Germany}%
\affiliation{J\"ulich Aachen Research Alliance (JARA), Fundamental of Future Information Technology, 52425 J\"ulich, Germany}%
\affiliation{Institute of Physics II, Universität zu Köln, 50937 Köln, Germany}%

\author{E. Il'ichev}%
 \affiliation{ Leibniz Institute of Photonic Technology, 07702 Jena, Germany}%

\date{\today}

\begin{abstract}
Due to their high kinetic inductance, highly disordered superconducting thin films are a potential hardware for the realization of compact, low-noise elements in cryoelectronic applications. However, high disorder typically results in structural defects that cause spatial inhomogeneity of the superconducting order parameter, thereby impairing the film's properties. Here, we employ scanning tunneling microscopy to demonstrate that NbN thin films fabricated by plasma-enhanced atomic layer deposition (PE-ALD) exhibit unusual spatial homogeneity, even at thicknesses approaching the superconductor-insulator transition. 
Tunneling spectra acquired across the sample surface show only small variations of the order parameter with a standard deviation of $2-3\%$, on length scales that significantly exceed the film's grain size.
At the same time, the films achieve a relatively high sheet resistance (up to 1400 Ohm) and, consequently, a high sheet kinetic inductance (up to approximately 200 pH), making them well-suited for applications.
\end{abstract}

\maketitle

\section{\label{sec:level1}Introduction}

Superconducting films that exhibit relatively high resistance in their normal state typically have high kinetic inductance, since \cite{tinkham}
\begin{equation}\label{eq:kin}
       L_\square = hR_\square/(2\pi^2\Delta).
\end{equation}

\noindent Here, $h$ is Planck’s constant, $\Delta$ is the superconducting order parameter, and $L_\square$, $R_\square$ are the sheet inductance and sheet resistance in the normal state, respectively.
The property of high kinetic inductance ($L_\square\gtrsim10\,\rm pH$) is relevant for various applications, including superinductors \cite{Niepce19}, efficient filters \cite{Shaikhaidarov22}, and parametric amplifiers \cite{Kern23}. 
Recently, niobium nitride (NbN) thin films with high kinetic inductance were further used to take an important step towards a quantum current standard\cite{Shaikhaidarov24}.  
For thin films, a high resistance in the normal state is often the result of disorder, which can be controlled, for instance, by changing the film thickness. Decreasing the film thickness generally causes an increase in granularity, which in turn causes more variations of the superconducting order parameter, leading to a broader superconducting transition and a lower critical temperature $T_\mathrm{c}$ \cite{Deutscher73}.

We distinguish two main scenarios to describe the properties of thin films with decreasing thickness. The first scenario is the so-called granular disorder, in which the film is considered as a random Josephson array \cite{grunhaupt2019granular}. In this scenario, the grains of the film are weakly coupled. Therefore, the superconducting properties are determined by the Josephson coupling between the grains. These films exhibit spatial inhomogeneities in the order parameter, and their critical current is significantly lower than the depairing current of the film material \cite{White86}. In the second scenario, the film grains are much better coupled and the resulting films are typically referred to as homogeneously disordered  \cite{Dynes86}. Their superconducting properties are similar to those of dirty superconductors, demonstrating relatively high critical current densities, close to the theoretical depairing current limit. In addition, homogeneously disordered films have a smaller spread of parameters and are subject to fewer fluctuations of the order parameter. Thus, homogeneously disordered films generally are more reproducible than granular films and preferable for applications. Nonetheless, disorder-induced variations in the order parameter of such films degrade their performance in applications. 

High-quality sputtered and ALD-grown superconducting NbN films are described by the homogeneously disordered scenario. NbN films exhibit a critical temperature of $T_\mathrm{c}\sim15\rm\,K$ and a relatively large superconducting order parameter, which reduces unwanted quasiparticle excitations. Because of their unique properties, NbN films have been successfully used in the above-noted applications \cite{Niepce19, Shaikhaidarov22, Kern23, Shaikhaidarov24}, including extremely sensitive photon detectors \cite{Gol'tsman01, Natarajan12}.

NbN films are usually fabricated by reactive magnetron sputtering of niobium in a nitrogen atmosphere. However, this manufacturing process lacks spatial homogeneity of film parameters due to a resulting high graininess of the film structure.
In contrast, PE-ALD of NbN provides slow and uniform wafer-scale film growth, as well as precise control of the deposited film thickness \cite{Ziegler13, Linzen17, Knehr21}. This is the result of a sequence of repeated deposition cycles, where each cycle includes precursor deposition and a hydrogen plasma reduction step, separated by intermediate purge processes of the vacuum chamber, resulting in a small integral growth rate. However, due to the chemical nature of the deposition process, oxygen, hydrogen and carbon impurities  are introduced into the resulting films \cite{Ziegler17}, causing them to exhibit a relatively high resistance in the normal state compared to the resistance of sputtered films. In addition, $T_\mathrm{c}$ for ALD-grown "bulk" films (when $d \gg \xi$, where $\xi \approx 5\rm\,nm$ \cite{Semenov09} is the coherence length) is about 14\,K \cite{Ziegler17}, which is lower than that of sputtered films \cite{Semenov09, Shy73}. This, in turn, results in a lower critical current density for ALD-deposited films compared to sputtered films at liquid helium temperature. Detailed information on the comparison of the properties of ALD-deposited and sputtered films is presented in Ref. \onlinecite{Knehr2019}. In spite of the slightly suppressed critical current density and critical temperature, the ALD-deposited films have demonstrated their potential for applications in superconducting quantum circuits. 
Due to their relatively high normal-state resistance, a novel type of superconducting weak link - the coherent quantum phase slip -  could be realized using ALD-grown NbN films \cite{Graaf2018}. The corresponding high kinetic inductance is further used as an effective protection of delicate quantum circuits from the environment  \cite{Shaikhaidarov22}, as well as for an amplification of weak microwave signals \cite{Kern23}. However, to achieve high values of kinetic inductance and normal state resistance the film thickness needs to be just above the superconductor-insulator transition. For such films, fluctuations and a non-uniform distribution of the superconducting order parameter are expected, which inevitably cause uncontrolled fluctuations in corresponding quantum circuits  \cite{Fegel10}.

In this study, we fabricated a set of NbN thin film samples with thicknesses $d$ ranging from 50\,nm to 4\,nm on thermally oxidized silicon as well as sapphire substrates. 
Our samples were grown using a (tert-butylimido)-tris (diethylamido)-niobium precursor (TBTDEN) and a substrate temperature of 380\,°C, resulting in a growth rate of 0.47\,\r{A} per cycle, and 0.26\,\r{A} per minute. 
X-ray diffraction (XRD) analysis of the films reveals a polycrystalline structure of the cubic superconducting $\sigma$-NbN phase \cite{Linzen17}. The in-plane orientation of the NbN film grains does not correlate with the axes of the single-crystal silicon or sapphire substrates, i.e., the PE-ALD process leads to non-epitaxial film growth. Using XRD data, we estimated the average size of the NbN grains to be approximately 10\,nm for the example of a 40\,nm thick film. Thus, the structure of films obtained by the PE-ALD method can be characterized as nanocrystalline.

\begin{figure}
\includegraphics[width=8.0cm]{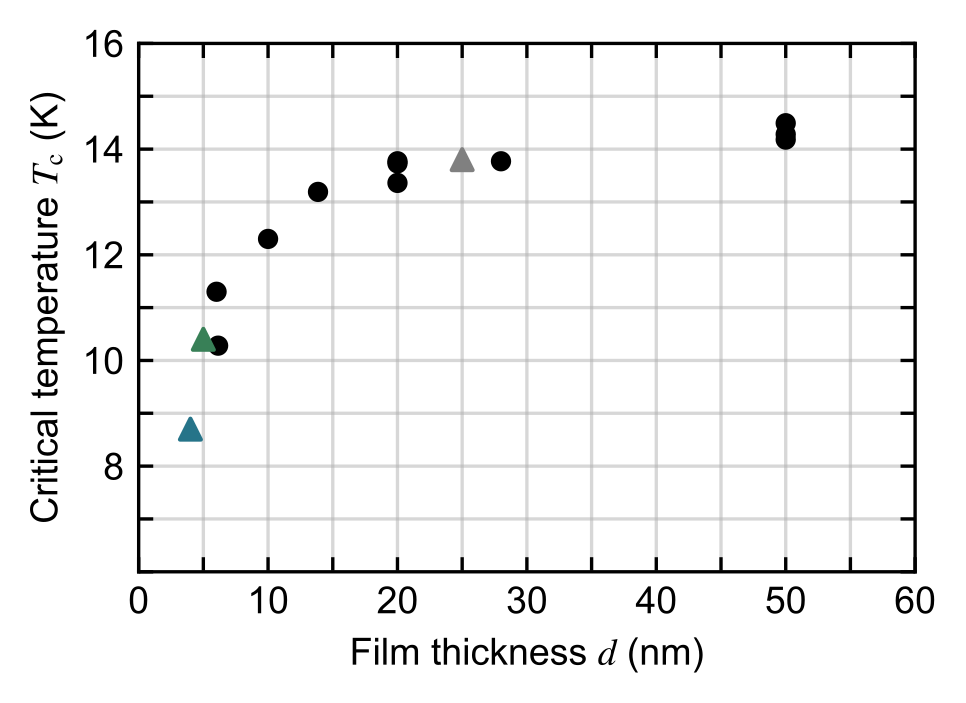} 
\caption{\label{fig:Tc_d_B} Critical temperature of NbN grown by PE-ALD on silicon substrates as a function of the film thickness. The colored triangles mark the data points of the 4\,nm, 5\,nm and 25\,nm films investigated in this study.}
\end{figure}

Standard temperature-dependent dc four-probe measurements were carried out by applying a probe current of 3\,\textmu A to unstructured films as well as to 10\,\textmu m\,x\,300\,\textmu m bridge structures, and the resulting $T_\mathrm{c}$ as a function of film thickness is shown in Fig.~\ref{fig:Tc_d_B}.
As the thickness decreases, $T_\mathrm{c}$ declines nonlinearly for film thicknesses below approximately 15\,nm, from approximately $14\rm\,K$ to below $9\rm\,K$ at 4\,nm.
In the following, a more detailed study was conducted on three sample films with thicknesses of 25\,nm, 5\,nm, and 4\,nm. The 25\,nm sample corresponds to the "bulk" material properties ($d \gg \xi$). For the thinner films (5\,nm and 4\,nm) $d \approx \xi$ and a quasi-two-dimensional character is expected. Indeed, the normal state sheet resistance of the thinner films increased significantly. At the same time, $T_\mathrm{c}$ decreased by only $\approx$ 30$ \%$ compared to the 25\,nm film.
For the 5\,nm and 4\,nm films, which have thicknesses close to the critical thickness $d_\mathrm{c}\approx3\,\rm nm$ of the superconductor-insulator transition in ALD-grown NbN \cite{Linzen17}, a strong spatial inhomogeneity of $\Delta$ can be expected due to disorder in the films \cite{Fegel10}. In addition, deviations from the BCS scenario of superconductivity are possible, such as a suppression of coherence peaks in the superconducting density of states \cite{Noat13}. To characterize these film properties, which are not accessible by standard transport measurements, we employ scanning tunneling microscopy.

Scanning tunneling measurements were performed in ultra-high vacuum (UHV), at a base temperature of 1.5\,K (for the 25\,nm film) and 1.2\,K (for the 5\,nm and 4\,nm film), respectively.
To minimize surface degradation of the grown films during transport from the ALD chamber to the STM, the samples were sealed \textit{in situ} under dry nitrogen atmosphere using a glovebox attached to the load lock of the ALD system. Prior to loading the samples into the preparation chamber of the STM, the samples were exposed to air for several minutes to mount them onto standard STM sample plates. After introduction into the UHV, the films were degassed for 20\,minutes at 550\,K to remove surface adsorbates that may result from the brief air exposure.
After cooling to room temperature, the samples were transferred to the STM.
For the tunneling experiments, PtIr tips were prepared on a clean Ag(111) single crystal by indenting the tip into the crystal surface until the tip shows a flat density of states. STM images of the NbN films were obtained by applying a bias voltage $V=0.8\,\rm V$  to the sample and maintaining a tunneling current of 300\,pA.
\begin{table*}
    \centering
    \begin{ruledtabular}
    \begin{tabular}{ccccccccc}
Deposition & Substrate & Deposition $T$ ($\,^\circ$C) & $d$\rm\,(nm) & $T_\mathrm{c}\rm\,(K)$ & $R_\square\rm\,(\Omega)$ & $\Delta\rm\,(meV)$ & $L_\square\rm\,(pH)$ & $\Delta/k_\mathrm{B} T_\mathrm{c}$\\
\hline
ALD &
        silicon & 380 & 25  & 
        $13.8 $ & 80 &  2.11 &  7.94 &1.77\\
ALD &
        sapphire & 380 & 5  &
        $10.4 $ & 900 &  2.02 & 93.4 &2.25\\
ALD &
        silicon & 380 & 4 &
        $8.7 $ &  1400 &
        1.60 & 183 & 2.13 \\

\hline
Sputtered (Ref. \onlinecite{Noat13}) &  sapphire  & 750 & 15 &
        $15.0 $ &
          $50$ &   2.85 & 3.68 & 2.20\\

Sputtered (Ref. \onlinecite{Noat13})&  sapphire & 750 & 8 &
        $14.5 $ &
          $100$ & 2.7 & 7.76 & 2.16\\

Sputtered (Ref. \onlinecite{Noat13})& sapphire & 750 & 4 &
        $13.3 $ &
          $140$ &  2.4 & 12.2 & 2.09\\
    \end{tabular}
    \end{ruledtabular}
    \caption{Comparison of parameters of PE-ALD grown NbN thin films from this work with sputter deposited films from Ref. \onlinecite{Noat13}. The "Deposition $T$" refers to the substrate temperature during deposition. Values for the sheet kinetic inductance $L_\square$ have been calculated using Eq. \ref{eq:kin}.}
    \label{tab:my_label}
\end{table*}
For scanning tunneling spectroscopy, a lock-in detection scheme was employed to acquire the differential tunneling conductance d$I$/d$V$ at a modulation frequency of 819\,Hz and a peak-to-peak amplitude of 70\,\textmu V, at a setpoint current of 1\,nA.

\begin{figure}[htp!]
\includegraphics[width=1\columnwidth]{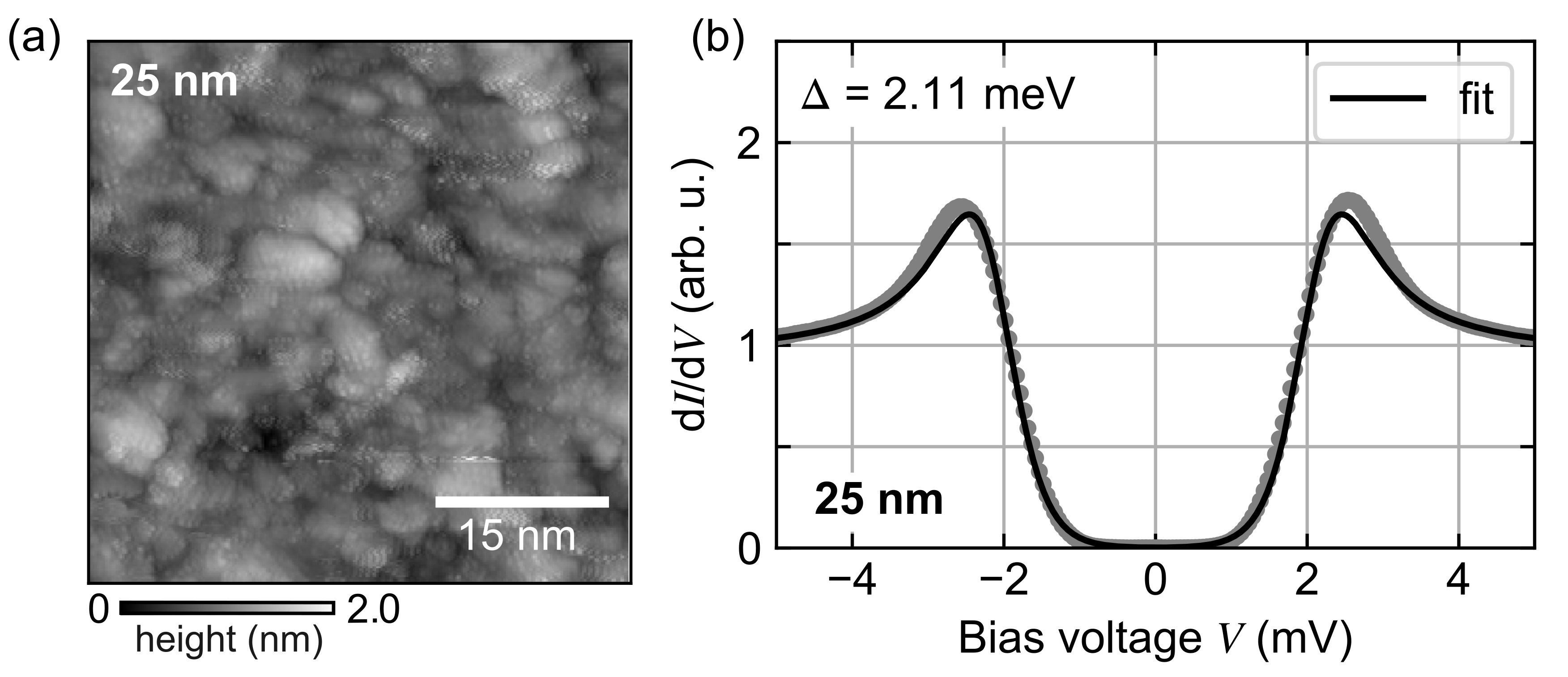}%
\caption{\label{fig:NbN_BCS_gap_fit_25nm} LT-STM analysis of a 25\,nm NbN film on silicon. (a) STM topography and (b) normalized tunneling spectrum averaged over 50 spectra taken along a 70\,nm line. The shown BCS fit (Eq. \ref{didv_fit} and \ref{bcs_dos}) yields an order parameter $\Delta=2.11\rm\,meV$ and an effective tip temperature $T_\mathrm{tip}=3.03\rm\,K$.}
\end{figure}

Fig.~\ref{fig:NbN_BCS_gap_fit_25nm} (a) depicts the STM topography of the 25\,nm film, showing characteristic grain diameters ranging from 3\,nm to 10\,nm.
Fig.~\ref{fig:NbN_BCS_gap_fit_25nm} (b) presents a spatially averaged tunneling conductance spectrum obtained by averaging 50 spectra measured along the 70\,nm diagonal of the STM image shown in Fig.~\ref{fig:NbN_BCS_gap_fit_25nm} (a). The original 50 spectra are plotted in Fig. S1 in the supplementary information. 
All spectra have been divided by a linear background fitted to the DOS outside the superconducting gap.
The order parameter $\Delta$ is extracted from the observed pairing gap using a simplified relation between the tunneling conductance ${dI}/{dV}$ and the sample DOS, which assumes a constant tunneling matrix element and tip DOS \cite{bardeen1961tunnelling,voigtlaender2015,esat2023determining}
\begin{equation} \label{didv_fit}
dI/dV
\propto \int_{-\infty}^{\infty}
\, dE \,
\left[ \frac{d}{dV} \, f_\mathrm{tip}(E)\right]
n_\mathrm{sample}(E+eV),
\end{equation}

\noindent where $I$ is the tunneling current, $V$ is the bias voltage, $n_\mathrm{sample}$ is the DOS of the superconducting sample, $e$ is the elementary charge, $E$ is the energy, and $f_\mathrm{tip}(E)=1/[1 + \exp(E/k_\mathrm{B} T_\mathrm{tip})$ is the Fermi-Dirac distribution function of the tunneling tip.

We treat NbN as a conventional $s$-wave superconductor. Therefore, the sample DOS can be described by the well-known expression from the Bardeen-Cooper-Schrieffer (BCS) theory \cite{tinkham}
\begin{equation} \label{bcs_dos}
n_\mathrm{sample}(E) =
n_0 \, \mathrm{Re} \left[ \frac{1}{\sqrt{E^2 - \Delta^2}} \right],
\end{equation}

\noindent where $n_0$ is the normal state DOS.
Using Eqs.~\ref{didv_fit} and \ref{bcs_dos} we fit the tunneling spectra by varying the fit parameters $\Delta$ and $T_\mathrm{tip}$.
A fit of the 25\,nm film spectrum is shown in Fig.~\ref{fig:NbN_BCS_gap_fit_25nm}(b) and describes the data well, resulting in $\Delta=2.11\,\rm meV$ and $T_\mathrm{tip}=3.03\rm\,K$.
We note that a higher tip temperature compared to the STM base temperature is typically the result of suboptimal tip thermalization or electronic noise. However, this does not affect the extracted values of $\Delta$ in our study.
Nonetheless, we optimized the experimental setup to lower $T_\mathrm{tip}$ for measurements on the thinner films.

\begin{figure}[htp!]
\includegraphics[width=1\columnwidth]{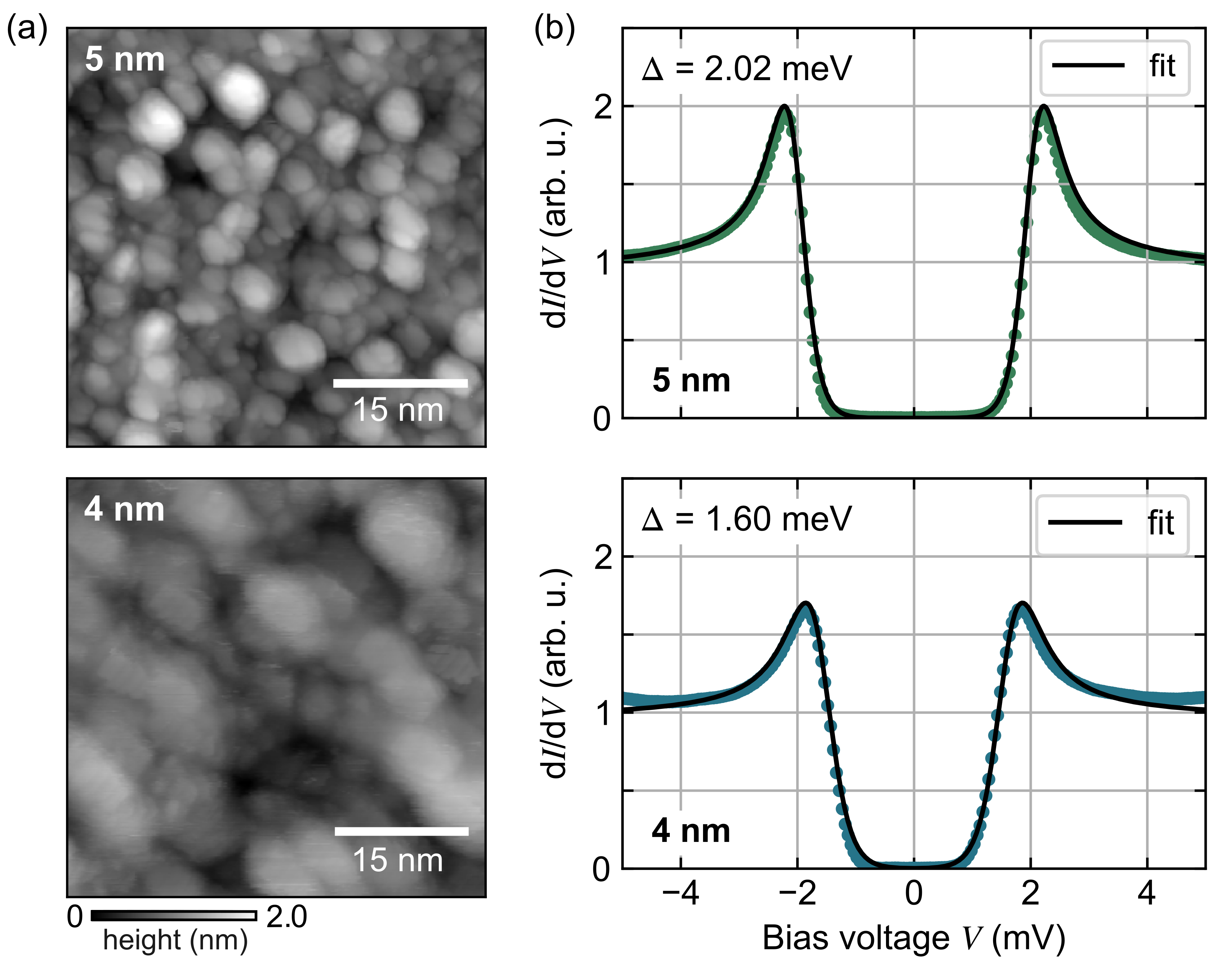}%
\caption{\label{fig:NbN_BCS_gap_fit} LT-STM analysis of a 5\,nm NbN film on sapphire and a 4\,nm NbN film on silicon. (a) STM topography and (b) normalized tunneling spectrum averaged over 200 spectra taken along a 424\,nm line for each of the two films. The shown BCS fits yield order parameters $\Delta=2.02\rm\,meV$ and $1.60\rm\,meV$ and effective tip temperatures $T_\mathrm{tip}=1.73\rm\,K$ and $2.27\rm\,K$ for the 5\,nm and 4\,nm film, respectively.}
\end{figure}

\begin{figure}[htp!]
\includegraphics[width=1\columnwidth]{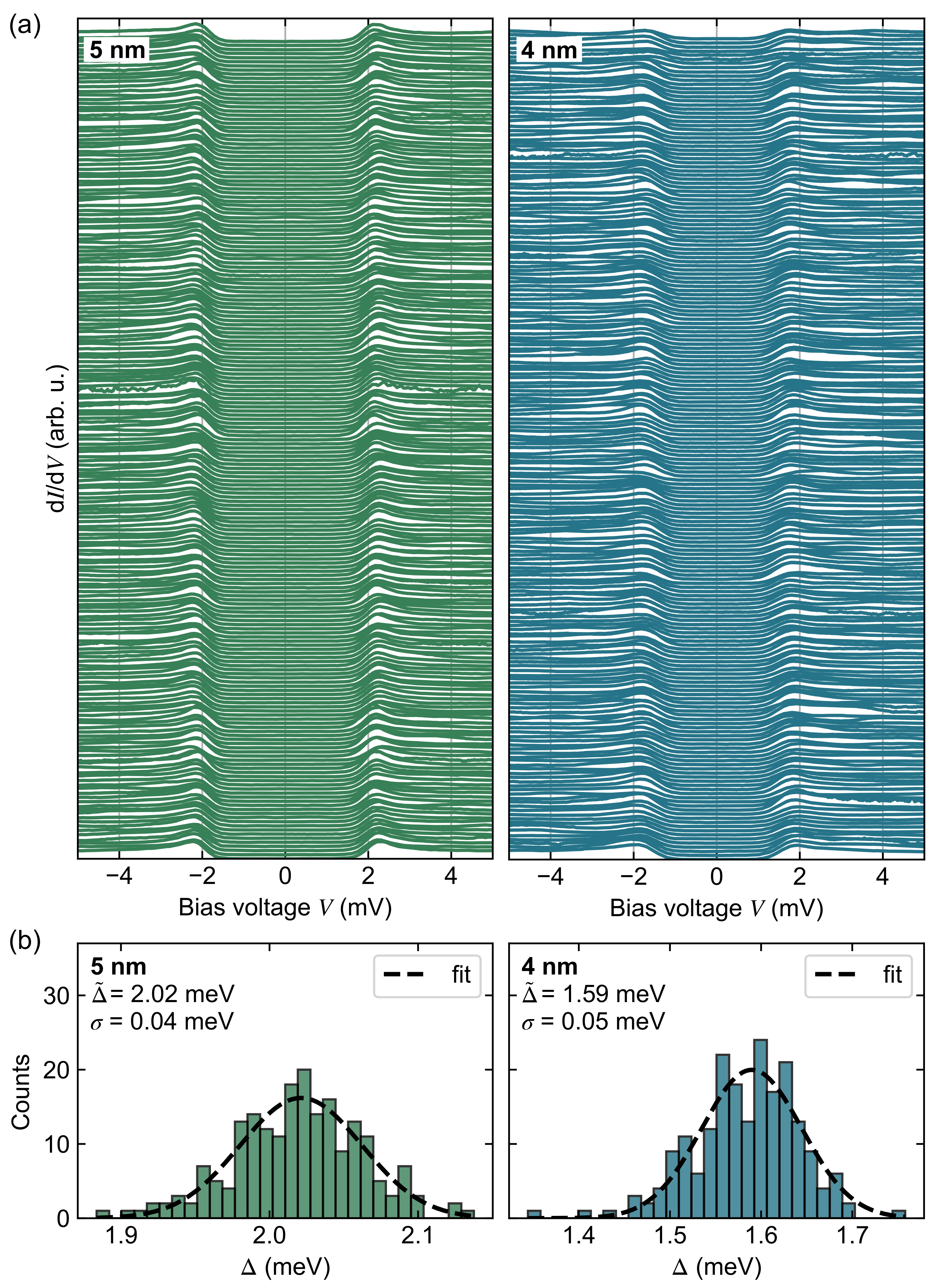}%
\caption{\label{fig:NbN_gap_statistics} Spatial homogeneity of the superconducting order parameter.
(a) Waterfall plot of the 200 tunneling spectra measured along a 424\,nm line across the 5\,nm NbN film on sapphire and 4\,nm NbN film on silicon. Spectra are offset for visual clarity.
(b) Histogram of the order parameters for both films obtained by BCS fits to each of the individual spectra shown in (a).
The mean value $\tilde\Delta$ and the standard deviation $\sigma$ of a Gaussian fit (Eq. \ref{gaussian_fit}) are indicated.}
\end{figure}

STM topographies and averaged STS spectra including corresponding BCS fits on the 5\,nm and 4\,nm films were obtained in a similar manner to the 25\,nm film and are shown in Fig.~\ref{fig:NbN_BCS_gap_fit}. For the 5\,nm and 4\,nm films we averaged 200 spectra measured along the 424\,nm diagonal of a 300\,x\,300\,nm$^2$ image. The original 200 spectra are displayed in Fig.~\ref{fig:NbN_gap_statistics} (a). 
The 300\,x\,300\,nm$^2$ scan size was chosen to analyze the spatial homogeneity across the film while still being able to resolve individual grains within reasonable measurement times ($\sim 5\rm\,h$). In addition, this length scale is commonly found in critical circuit elements, like nanowires\cite{Shaikhaidarov22}, which often have widths of $\sim100\,$nm. To further check that the superconducting properties are homogeneous across the whole film, we repeated d$I$/d$V$ measurements in different spots on the 10\,x\,10\,mm$^2$ samples (Fig.\,S1) and include a d$I$/d$V$ map of a 50\,x\,50\,nm$^2$ area taken on the 4\,nm sample (Fig.\,S2).
Coming back to Fig.~\ref{fig:NbN_BCS_gap_fit}, the corresponding fits of Eq.~\ref{didv_fit} again show excellent agreement with the data, resulting in $\Delta=2.02\rm\,meV$ and $1.59\rm\,meV$ for the 5\,nm and 4\,nm film, respectively.
The superconducting properties of the investigated films are summarized in the top part of Table~\ref{tab:my_label}.
Comparing our results to NbN films produced by reactive magnetron sputtering (bottom part of Table~\ref{tab:my_label}), it is evident that $T_\mathrm{c}$ and $\Delta$ values of NbN thin films produced by PE-ALD are significantly lower, resulting in larger $L_\square$, as calculated from Eq.~\ref{eq:kin}.
In addition, the NbN thin films obtained by PE-ALD exhibit an improved spatial homogeneity of the order parameter $\Delta$, which is advantageous for device fabrication.
Fig.~\ref{fig:NbN_gap_statistics} (a) displays the individual spectra of the 200 spectra measured on the 5\,nm and 4\,nm thick films.
For a statistical analysis, we plot a histogram of the order parameters in Fig.~\ref{fig:NbN_gap_statistics} (b), which we obtain by fitting each individual spectrum with Eq. \ref{didv_fit}.
Fitting the histograms with a Gaussian function
\begin{equation} \label{gaussian_fit}
g(\Delta) \propto \exp\left( -\frac{1}{2} \frac{(\Delta - \tilde\Delta) ^2} {\sigma ^2}  \right),
\end{equation}
we extract the mean value $\tilde\Delta=2.02\rm\,meV$ and the standard deviation $\sigma=0.04\rm\,meV$ for the 5\,nm film and
 $\tilde\Delta=1.59\rm\,meV$ and $\sigma=0.05\rm\,meV$ for the 4\,nm film, as indicated in Fig. \ref{fig:NbN_gap_statistics}.

The small $\sigma$ of only $2-3\%$ of $\tilde\Delta$, and the fact that there are only a few outliers outside of the Gaussian distribution indicates that $\Delta$ is very uniform across the film.
We note that the spectra were measured over a distance much larger than the characteristic grain sizes and that the film thicknesses are close to the critical thickness. Therefore, some structural defects are expected.
However, if they exist, they do not affect the order parameter significantly. 
For films obtained by reactive magnetron sputtering, the situation is quite different: Although only "tiny spatial inhomogeneities of the SC gap" are observed for 8\,nm thick samples \cite{Noat13}, structural defects are clearly visible at similar scanning sizes (see Fig.~2 in Ref. \onlinecite{Noat13}). For their 4\,nm thick sample, spatial inhomogeneities become more noticeable and are characterized by two spatial length scales. The first scale is of the order of several nanometers, probably due to grain boundaries, and the second scale is of several tens of nanometers, due to the variation in film thickness \cite{Noat13}. It is worth noting that the superconductor-insulator transition for these sputtered films occurs at a lower thickness of approximately  $d_\mathrm{c} \approx 2$\,nm compared to our ALD-grown films.

Lastly, we note that the approximate theoretical BCS parameter $(\Delta/k_\mathrm{B}T_{c})_\mathrm{BCS} = 1.76$ (Ref. \onlinecite{tinkham}) is only observed in our thickest studied ALD film. For all other films, $\Delta/k_\mathrm{B} T_\mathrm{c}\approx 2.14\pm0.05$ is significantly higher than the BCS value and does not depend on neither the film thickness nor the growth method.
A summary of the studied films is given in Table \ref{tab:my_label}.
The relatively high values of $T_\mathrm{c} \approx 10$ K and $L_\square \approx 100-200$ pH (estimated from Eq.~\ref{eq:kin}) are well-suited for applications, as discussed in the introduction. 

In conclusion, we demonstrated that NbN thin films obtained by PE-ALD exhibit slightly lower superconducting parameters $\Delta$ and $T_{\rm c}$ compared to films fabricated by reactive magnetron sputtering. However, even at thicknesses close to the critical thickness, our scanning probe characterization shows that films grown by PE-ALD exhibit unusual spatial homogeneity of the order parameter at the nanoscale. We attribute this to the well-controlled thickness in ALD processes and small grain size of the films.
Scanning tunneling microscope measurements thus complement studies of common transport characteristics like sheet resistance\cite{Lennon2023} and sheet kinetic inductance with microscopic measurements of the order parameter. These homogeneously disordered films are well-suited for various superconducting devices, such as those relying on nanometer scale structures, e.g., in SNSPDs.

This work was partially supported by the German Federal Ministry of Research, Technology and Space (BMFTR) under Grant No. 13N16152/QSolid and "NbNanoQ", Grant No. 13N17121.
F.L. acknowledges funding from the Emmy Noether Programme of the DFG (Project No. 511561801) and Germany's Excellence Strategy-Cluster of Excellence Matter and Light for Quantum Computing (ML4Q) through an Independence Grant.
F.L. and F.S.T. acknowledge funding from the Bavarian Ministry of Economic Affairs, Regional Development and Energy within Bavaria's High-Tech Agenda Project “Bausteine für das Quantencomputing auf Basis topologischer Materialien mit experimentellen und theoretischen Ansätzen.”\\

\section*{Author declarations}
\subsection*{Conflict of Interest}
The authors have no conflicts to disclose.

\subsection*{Author contributions}
S.L., F.S.T., F.L. and E.I. conceived the project.
S.L., M.Z., G.O., R.S. grew the samples and performed the transport measurements.
J.L. conducted the STM measurements and analyzed the corresponding data.
J.L., S.L., F.L. and E.I. wrote the manuscript.

\section*{Data Availability}
The data that support the findings of this study are available from the corresponding author upon reasonable request.

\section*{References}
\bibliography{Referenzen_Sven_review.bib}

\end{document}